\documentclass[final,3p,times,twocolumn]{elsarticle}

\usepackage{amssymb}

\journal{Physics Letters B}
\def\frazione{\xi}
\def\costanteFlusso{F}

\begin{document}

\begin{frontmatter}



\title{Testing neutrino decay scenarios with IceCube data}

\author{G. Pagliaroli}
\address{INFN, Laboratori Nazionali del Gran Sasso, Assergi (AQ), Italy}
\author{A. Palladino}
\address{Gran Sasso Science Institute, L'Aquila (AQ), Italy}
\author{F. Vissani}
\address{INFN, Laboratori Nazionali del Gran Sasso, Assergi (AQ), Italy}
\address{Gran Sasso Science Institute, L'Aquila (AQ), Italy}
\author{F.L. Villante}
\address{L'Aquila University, Physics and Chemistry Department, L'Aquila, Italy}
\address{INFN, Laboratori Nazionali del Gran Sasso, Assergi (AQ),  Italy}


\begin{abstract}
We test the hypothesis of non-radiative neutrinos decay using 
the latest IceCube data. Namely, we calculate the 
track-to-shower ratio expected in IceCube for 
the normal and inverted neutrino mass hierarchy 
taking into account the uncertainties in neutrino oscillation parameters. 
We show that the subset of data with energy above 60 TeV 
actually excludes the possibility of a neutrinos decay 
at the 1 sigma level of significance for both neutrino mass hierarchies. 
\end{abstract}


\end{frontmatter}


\section{Introduction}
\label{sec:1}
The neutrinos decay scenario has been early proposed
as an explanation of the solar neutrinos
problem\cite{Bahcall:1972my,Pakvasa:1972gz}. 
Although the neutrinos decay through radiative processes is
well constrained \cite{Agashe:2014kda}, the lifetime 
limits for non-radiative channels are very weak. 
Generally researchers focused their attention to
non-radiative processes of the kind
$\nu'\rightarrow \nu+X$, where neutrinos decay into possibly 
detectable neutrinos (or antineutrinos) plus truly invisible
particles, $X$, e.g., light scalar or pseudoscalar bosons.
As an example, the neutrino decay can take place through  
the coupling of the neutrino to a very light 
or massless particle, such as a Majoron, which is also responsible 
for spontaneous neutrino mass generation \cite{Chikashige:1980ui,Gelmini:1980re}.

In general, considering a neutrinos source, 
decay could deplete the flux of neutrinos by the factor,
$$
\exp\left( - \frac{T}{E}  \times \frac{m_i}{\tau_i} \right)
$$
where $m_i$ and $\tau_i$ are the masses and lifetime of the $i$-th neutrino, subject to decay,
$E$ is their energy and $T$ is the time since production.
As a consequence, the sensitivity to the unknown 
ratio $\tau/m$ changes for different distances and
energies.
Actually the strongest reliable limit for non-radiative processes and 
for hierarchical masses, namely $\tau/m>1.1\cdot 10^{-3}$ s/eV,
is obtained by the non-observation of solar electron 
antineutrinos in Kamland \cite{Eguchi:2003gg}. 
This limit can be in principle improved by several order of magnitude 
by using the high energy cosmic neutrinos observed in detectors such 
as IceCube \cite{Beacom:2002vi}. 

This intriguing possibility became
more interesting on the light of the recent search for 
High Energy Starting Events (HESE) in IceCube detector that
provided the first evidence for high-energy cosmic 
neutrinos\cite{IceCube3years,IceCube1TeV,IceCubeScience}. 
In three year of data taking \cite{IceCube3years}, 
37 events with deposited energies above 30 TeV 
were observed, relative to an expected
background of $8.4 \pm 4.2$ cosmic ray muon events and
$6.6\pm $5.9 atmospheric neutrinos. 


Recently, Palladino {\it et al.}\cite{Palladino:2015zua} have 
discussed the compatibility of these data with the hypothesis 
of cosmic origin.  In order to reduce the background contamination, 
authors have considered the subset of events with deposited 
energy above 60 TeV. Using the additional information provided 
by muon neutrinos passing through the Earth, they show that the 
observed track-to-shower ratio matches expectations for neutrinos of cosmic origin.

\begin{figure*}[ht]
\includegraphics[width=1.\textwidth,angle=0]{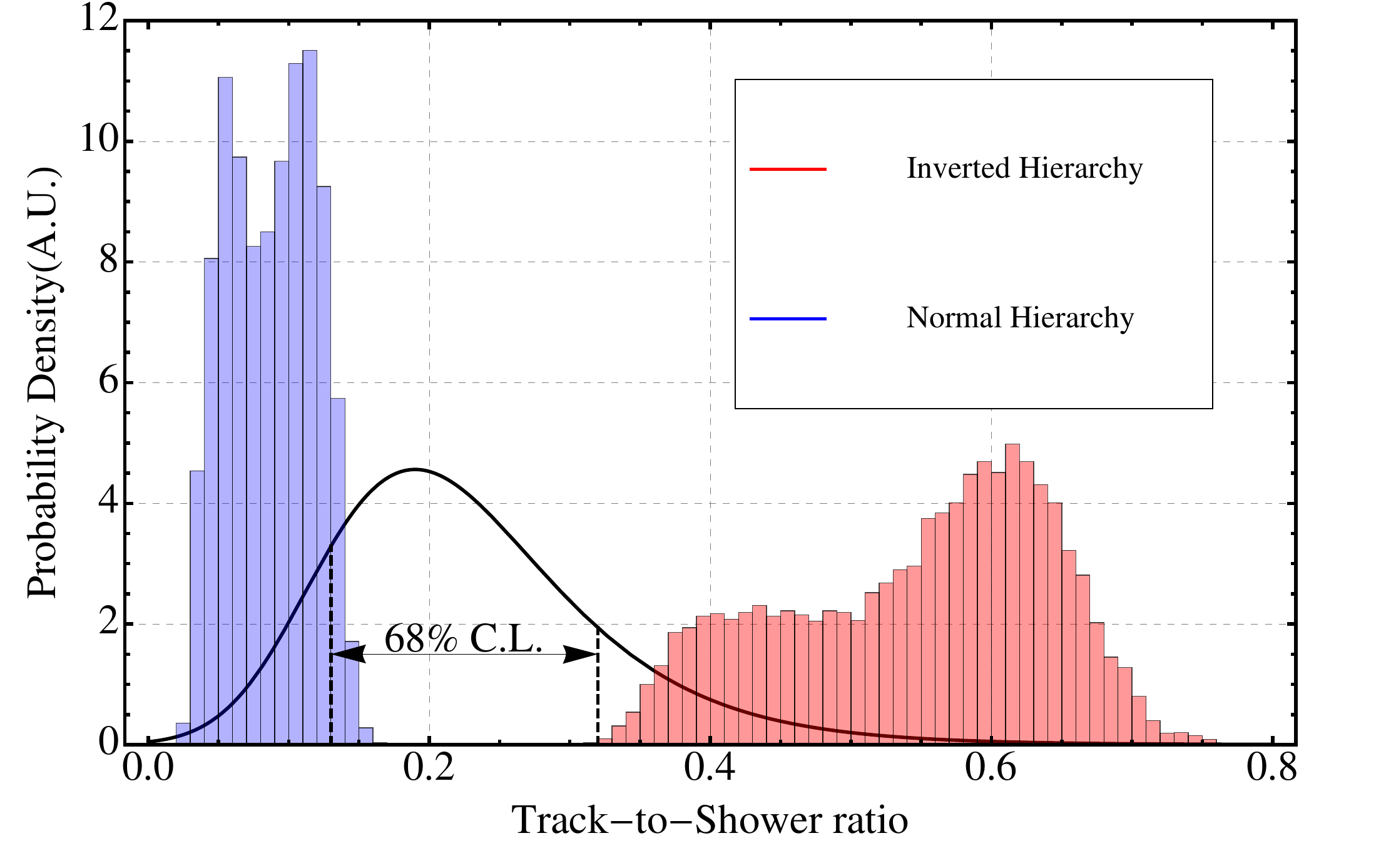}
\vspace{-7mm} \caption{\em\protect\small  
Expected track-to-shower ratio for decaying cosmic neutrinos. The distributions show the 
effect of uncertainties in the neutrino oscillation parameters. 
The left (resp.\ right) distribution is obtained for normal (resp.\
inverse) hierarchy. The black line shows the normalized likelihood
function of the IceCube data-set discussed in the text.}
\label{fig1}
\end{figure*} 

In this Letter, we discuss the impact of the non-radiative 
neutrino decay scenario on the expected IceCube signal 
in term of the track-to-shower ratio. 
More precisely, considering the actual 
uncertainties on oscillation parameters, mass hierarchy
and production mechanisms; we use the observed events
to constrain the hypothesis of neutrinos decay.
As a remarkable result, we show that the subset 
of data with energy above 60 TeV excludes 
the possibility of a neutrinos decay 
at the 1 sigma level of significance for 
both neutrino mass hierarchies. 

\section{Complete Decay Scenario}
\label{theory}

To maximize the effect of neutrino decay, 
we consider the phenomelogy of a generic neutrinos decay 
and we assume that decays are always complete ( i.e., that the exponential
factors vanish). 
This last assumption is reasonable for Cosmic Sources 
considering that $L\sim 1\mbox{ Gpc}$ and $E\sim100\mbox{ TeV})$, 
namely
\begin{equation}
\frac{T/E}{\mbox{s/eV}} =10^{3} 
\end{equation}
where we used $L=c T$. We call this hypothesis {\em Complete Decay Scenario}.
In this case, the flux of neutrinos of flavor $\ell$ 
expected to the Earth is: 
$$
\Phi_\ell
= \sum_{\ell'} P_{\ell\ell'} \, \Phi_{\ell'}^0 
\;\;\;\;\;
\mbox{with} 
\;\;\;\;\;
P_{\ell\ell'}=\sum_{j=stable} |U_{\ell j}|^2 |\ U_{\ell' j}|^2,
$$
where $U$ is the neutrino mixing matrix and the sum only involves 
stable mass eigenstates.
Following \cite{Beacom:2002vi,Palladino:2015zua}, we introduce
the {\em flavor fractions at Earth} (i.e., in the detection point), 
defined as:
\begin{equation}\label{ciups}
\frazione_\ell\equiv \Phi_\ell/\Phi_{\rm tot}
\;\;\;\;\;
\mbox{with} 
\;\;\;\;\;
\Phi_{\rm tot} =\sum_{\ell=active}\Phi_\ell.
\end{equation}
In term of this quantity the relation between the flavor ratios at
Earth and the ones at source becomes:
\begin{equation}
\label{fractionnh}
\frazione_{\ell} =\frac{\sum_{j}|U_{\ell j}|^2 \sum_{\ell'} \ |U_{\ell' j}|^2
  \frazione_{\ell'}^0}{\sum_{j,\ell}|U_{\ell j}|^2 \sum_{\ell' } |U_{\ell'j}|^2\frazione_{\ell'}^0},
\end{equation}
where the indeces $j$ and $\ell'$ run over stable mass eigenstates and
active flavors, respectively. The expression in
Eq.(\ref{fractionnh}) also holds in the case where sterile 
neutrino states are included. 
Different production mechanisms generate a
different initial flavor content at the source $(\frazione^0_e :\frazione^0_{\mu}
:\frazione^0_{\tau})$. For example the most studied production
processes are:\\
$(1/3 : 2/3 : 0)$ for $\pi$ decay;\\
$(1/2 : 1/2 : 0)$ for  {\em charmed mesons} decay;\\ 
$(1 : 0 : 0)$ for $\beta$ decay of {\em neutrons};\\
$(0 : 1 : 0)$ for $\pi$ decay with {\em damped
  muons}.\\ 
For cosmic neutrinos, without neutrino decay processes, 
the final flavor ratios are expected to be
very near to $(\xi_e:\xi_{\mu}:\xi_{\tau}) \sim (1/3 : 1/3 : 1/3)$,
independently on the specific production mechanism \cite{Learned,athar,Others,Noi,cinese}.
On the other hand, turning on the "invisible" 
decay process discussed above, the flavor ratios at the Earth 
change dramatically \cite{Beacom:2002vi}.
In the case of only one stable mass eigenstate $j$ and for the
standard scenario with three actives neutrinos, the expression of Eq.(\ref{fractionnh})
becomes  
\begin{equation}
\label{simp}
\frazione_{\ell} =|U_{\ell j}|^2,
\end{equation}
where the sensitivity to the different production mechanisms has
disappeared. 
The simplified expression in Eq.(\ref{simp}) is due to the fact that
only one mass eigenstate is stable and all the others have 
had time to decay, so that the universe is populated only by neutrinos 
in the state of mass $j$ and obviously the content of flavor $\ell$
expected to the detector is $|U_{\ell j}|^2$.
  
Assuming normal mass hierarchy, the simplest case
is to consider that both $\nu_3$ and $\nu_2$ decay, so that 
only the lightest eigenstate $\nu_1$ survives. In this case the 
final flavor ratios are expected to be $(|U_{e 1}|^2:|U_{\mu
  1}|^2:|U_{\tau 1}|^2)\sim (0.68 : 0.21 : 0.11)$.
For inverted mass hierarchy, a similar expression holds but the
stable state in this case is $\nu_3$, i.e. the final flavor ratios 
are expected to be $(|U_{e 3}|^2:|U_{\mu  3}|^2:|U_{\tau 3}|^2)\sim (0.02 : 0.57 : 0.41)$.  
In order to evaluate these ratios, we used the best-fit values 
for the oscillation parameters reported in \cite{nufit}.

\section{Track-to-Shower Ratio}
 \label{observable}
To compare the expectations with the data collected by IceCube 
we need to convert the flavor neutrino ratio 
expected to the Earth in the corresponding observable quantity, i.e. the
track-to-shower ratio including the detector response.   
Indeed, by exploiting the event topology, the flavor 
discrimination is possible, in principle, in IceCube detector.
The HESE data collected by IceCube during $T=988$ days, 
encompass two different topologies: `shower' topology, 
that includes neutral current (NC) interactions of all neutrino flavors and 
charged current (CC) interactions of 
$\nu_{\mathrm{e}}$  and $\nu_\tau$; 
`track' topology produced by CC interactions of $\nu_\mu$.
Thus the track-to-shower ratio is the observable quantity that is most
directly related to the flavor ratio at the Earth as discussed in
several recent papers \cite{Palladino:2015zua,Mena,Mena2,Icenew,Ruiz2}.

In particular Palladino {\it et al.}\cite{Palladino:2015zua} described 
the way to calculate the number of showers $N_{\rm S}$ 
and tracks $N_{\rm T}$ in the IceCube detector for an 
isotropic flux $\Phi_{\ell}$ of neutrinos and antineutrinos described
as \begin{equation}
\Phi_\ell(E) =\frac{\costanteFlusso_\ell \cdot
  10^{-8}}{\mbox{cm$^2$ s sr GeV}} \cdot
\left(\frac{\mbox{GeV}}{E}\right)^\alpha 
\label{fluxes}
\end{equation} 
where the factors $\costanteFlusso_\ell$ are (non-negative) adimensional coefficients 
and  $\alpha$ is the spectral index. 
The number of showers and tracks can be obtained by
using the following equations:
\begin{eqnarray}
N_{\rm S} &=& T \left\{ \costanteFlusso_{e}\,  c_{e} +
  \costanteFlusso_{\tau} \,  c_{\tau} +\costanteFlusso_\mu \, c_\mu (1-p_{\rm T})\right\},\\
N_{\rm T} &=& T \costanteFlusso_\mu \, c_\mu p_{\rm T},
\label{NTandNS}
\end{eqnarray}
where $T$ is the time of exposure and the coefficients 
$c_l= 4\pi \, \int dE \; E^{-\alpha}  \,A_{\ell}(E)$ include 
the detector effective areas $A_{\ell}(E)$\cite{IceCubeScience} and
the energy dependence of the fluxes. The extra factor $p_{\rm T}$ in 
Eq.(\ref{NTandNS}) accounts for the fraction of the effective area 
$A_\mu(E)$ giving tracks, this fraction is about $\sim 0.8$ and 
mildly dependent on energy as discussed in \cite{Palladino:2015zua}.
So that, the following useful expression to obtain the track-to-shower ratio 
from the final neutrino flavor ratios can be obtained generally:
\begin{equation}
\frac{N_{\rm T}}{N_{\rm S}} = \frac{\frazione_\mu}{c_1 +  c_2 \, \frazione_\mu +
  c_3 \, \frazione_\tau},
\label{NToverNs}
\end{equation}
in which $c_j$ are given, with good approximation, by:
\small
\begin{eqnarray}
&c_1=\frac{c_e}{p_{\rm T} \cdot c_\mu}  \simeq 2.23+1.15(\alpha-2)+0.46(\alpha-2)^2
\nonumber  \\
&c_2=\frac{(1-p_{\rm T})c_\mu-c_e}{p_{\rm T}\cdot c_\mu} \simeq
-1.99-1.15(\alpha-2)-0.46 (\alpha-2)^2   \nonumber   \\
&c_3=\frac{c_\tau-c_e}{p_{\rm T} \cdot c_\mu}  \simeq -0.56-0.78(\alpha-2)-0.40
(\alpha-2)^2 \nonumber 
\end{eqnarray}
\normalsize
In the context of neutrino decay scenario described above and fixing the spectral
index $\alpha=2$, we have that for Normal mass hierarchy Eq.(\ref{NToverNs})
becomes: 
\begin{equation}
\frac{N_{\rm T}}{N_{\rm S}} = \frac{|U_{\mu
  j}|^2}{2.3 -  2.0 \, |U_{\mu
  j}|^2 -0.6 \, |U_{\tau j}|^2},
\end{equation}
where the index $j$ indicates the stable mass eigenstate.
For Normal Mass Hierarchy $j=1$ and substituting the 
best-fit values of the oscillation parameters we
obtain $(\frac{N_{\rm T}}{N_{\rm S}})^{NH} = 0.12$. 
For Inverted Mass Hierarchy $j=3$ and this fraction becomes 
$(\frac{N_{\rm T}}{N_{\rm S}})^{IH} = 0.62$. The impact of the spectral index of neutrino spectrum is not so dramatic, in fact if we consider a greater slope, like $\alpha=2.6$, these number become $(\frac{N_{\rm T}}{N_{\rm S}})^{NH} = 0.09$ and $(\frac{N_{\rm T}}{N_{\rm S}})^{IH} = 0.56$.  Instead, a much more important contribution is given by uncertainties on neutrino oscillations parameters, that allow these numbers to
fluctuate in a more large interval. To correctly account for these uncertainties 
we construct likelihood distributions of $\sin^2 \theta_{12} $,
$\sin^2 \theta_{13} $, $\sin^2 \theta_{23}$ and $\delta$ from the 
$\Delta\chi^2$ profiles given by \cite{nufit}. Namely, we assume
that the probability distributions of each parameter are provided by
${\mathcal L} = \exp{\left(-\Delta \chi^2/2\right)}$. Then, we combine
the various likelihood functions assuming negligible correlations
and we determine the probability distributions of $N_{\rm T}/N_{\rm S}$ 
by MonteCarlo extraction of the oscillation parameters.
The resulting Probability Density (PD) distributions are 
reported in Fig.(\ref{fig1}) with a color code blue (resp. red) for
Normal (resp. Inverted) Mass Hierachy.
 
The predicted PDs can be compared with the distribution of the
track-to-shower ratio preferred by the IceCube data.
In the first column of Tab. (\ref{datatable}) we report the events
observed by IceCube in the HESE data set \cite{IceCube3years} with a
deposited energy above 60 TeV. We devided the events by topology and
we includ, in brackets, the respective number of estimated background events
due to atmospheric neutrinos and muons. 

The number of tracks $N_{\rm T}$ and showers  $N_{\rm S}$ which have 
to be ascribed to extraterrestrial sources can be estimated from the 
Poisson likelihood functions:
\begin{equation}
\mathcal{L}(N_{\rm i}) \propto \lambda_{\rm i}^{n_{\rm i}} \times  e^{-\lambda_{\rm i}}  \nonumber
\end{equation}
where $\lambda_{\rm i}=N_{\rm i} + b_{\rm i}$ and the index ${\rm i}={\rm T,S}$ is used to
refer to track and shower events and $n_{\rm i}$ are the observed events.
We are assuming that the prompt atmospheric neutrinos give negligible
contributions, as it required by the spectral and arrival angles
distributions of IceCube events. 
\begin{table}[b]
\caption{\footnotesize{The total number of events, $n_{\rm i}$, with the
    expected background, $b_i$, seen at IceCube, in the energy range 
$60 \mbox{TeV} \leq E_{dep} \leq 3 \mbox{PeV}$.}}
\begin{tabular}{|c|c|c|}
\hline	
&$n_{\rm i}$ ($b_i$) 988 days &$n_{\rm i}$ ($b_i$) 1460 days \\
\hline
Track	  &  4 (2.1)  &     8 (3.1)   \\
\hline
 Shower   & 16 (0.7)  & 25 (1.0) \\
\hline
Total    & 20 (2.8)  & 33 (4.1) \\
\hline
\end{tabular}
\label{datatable}
\end{table}
A completely equivalent and
independent information can be obtained by the recently released
IceCube data on passing muons \cite{IceCubePassingMuons}, as already
done by Palladino et al. in \cite{Palladino:2015zua}.
In the assumption of 
$E^{-2}$ neutrino spectrum for the passing muons, the flux normalization corresponds to
 $F_\mu=1.01\pm 0.35$ and taking into account the equivalence between $F_\mu$ and $N_T$
expressed by the following equation (in which $T$ is the exposure time):
\begin{eqnarray} 
N_{\rm S} &=& (2.94\times  F_e + 0.33 \times  F_\mu+2.20\times  F_\tau) \cdot T  \nonumber \\
N_{\rm T} &=& 1.31\times  F_\mu \cdot T \nonumber
\label{NTandNsSimple}
\end{eqnarray}
we can include also this information in our analysis by constructing a combined likelihood, given by the product of the 2 Poisson likelihoods for
 $N_{\rm T}$ and $N_{\rm S}$ and of the Gaussian likelihood for $F_\mu$ (i.e. $N_T$).
Then we extract the bounds on the track-to.shower ratios of events
ascribed to cosmic neutrinos by marginalizing with respect 
to the total number of events. We obtain: 
\begin{equation}
\frac{N_{\rm T}}{N_{\rm S}}=0.18^{+0.13}_{-0.05}  \mbox{\ \ \ \ [988 days]} \nonumber
\label{1}
\end{equation}
where the error was obtained by integrating out symmetrically $(1 -{\rm CL})/2$ on both
sides of the $N_{\rm T}/N_{\rm S}$ distribution using a confidence level 
${\rm CL} =68.3\%$.

The normalized likelihood distribution of the HESE data is reported in
Fig.(\ref{fig1}) with a black line and the 1 sigma region is also
indicated to easily see its sovrapposition with the predicted PDs in the
neutrino decay scenario.

\section{Discussion and conclusion} 

Repeating the analysis with 4 years data, under the hypothesis that
the background is simply proportional to the exposure time, we found a new range given by:
\begin{equation}
\frac{N_{\rm T}}{N_{\rm S}}=0.20^{+0.10}_{-0.05}  \mbox{\ \ \ \ [4 years]} \nonumber
\label{2}
\end{equation}
We can notice that this interval is compatible with that found for the 3 year\cite{Palladino:2015zua}
also if the uncertainties are slightly smaller with new data;
the track-to-shower of 4 years is more favorable for pions decay at the source, respect to that obtained with 3 years data. 
In conclusion, it is important to remark that the non-radiative
neutrino decays are excluded at 
least at 1 sigma for both neutrino hierarchies and for both sets 
of observed data. Moreover these are general considerations and 
have a poor dependence from the slope of the spectrum of neutrinos.






\end{document}